\documentclass[pre,twocolumn,superscriptaddress]{revtex4-1}
\usepackage{amsmath,amssymb,hyperref,graphicx,subfigure,color}
\usepackage{epstopdf}

\begin{document}
\title{Interplay between finite resources and local defect in an asymmetric simple exclusion process}
\author{L. Jonathan Cook}
\affiliation{ Department of Physics and Engineering, Washington and Lee University, Lexington, VA 24450}
\author{J.\ J.\ Dong}
\affiliation{Department of Physics and Astronomy, Bucknell University, Lewisburg, PA 17837}
\author{Alexander LaFleur}
\affiliation{ Department of Physics and Engineering, Washington and Lee University, Lexington, VA 24450}

\begin{abstract}
When particle flux is regulated by multiple factors such as particle supply and varying transport rate, it is important to identify the respective dominant regimes. We extend the well-studied totally asymmetric simple exclusion model to investigate the interplay between a controlled entrance and a local defect site. The model  mimics cellular transport phenomena where there is typically a finite particle pool and non-uniform moving rates due to biochemical kinetics. Our simulations reveal regions where, despite an increasing particle supply, the current remains constant while particles redistribute in the system. Exploiting a domain wall approach with mean-field approximation, we provide a theoretical ground for our findings. The results in steady state current and density profiles provide quantitative insights into the regulation of the transcription and translation process in bacterial protein synthesis.  We investigate the totally asymmetric simple exclusion model with controlled entrance and a defect site in the bulk to mimic the finite particle pool and non-uniform moving rates in particle transport processes. 
\end{abstract}

\maketitle

\section{Introduction}

One of the paradigms in non-equilibrium statistical mechanics, the totally asymmetric simple exclusion process (TASEP) brings insights to various transport phenomena in stochastic systems.  Originally proposed in the context of protein synthesis \cite{MacDonald68,*MacDonald69} and pure mathematics \cite{Spitzer70}, TASEP now finds its versatility in biological transport \cite{MacDonald68,Shaw03,Chou04,Dong07b,Kolomeisky98,Lakatos03,Dong07,Chowdhury05}, traffic flow \cite{Chowdhury00,Popkov01,Ha02}, surface growth \cite{Kardar86} and much beyond \cite{Zia11,DKZ2012}.

The simple TASEP consists of particles moving uni-directionally along a one-dimensional lattice (at a site-dependent rate $\gamma_i$, typically unity for all lattice sites) with particles experiencing hard-core exclusion.  For periodic boundary conditions, the stationary distribution is trivial \cite{Spitzer70} but contains rich dynamics \cite{DeMasi85,*Kutner85,*Majumdar91,*Gwa92,*Derrida93b,*Kim95,*Golinelli05}.  With open boundary conditions where particles enter with rate $\alpha$ and exit with $\beta$, three phases emerge and the steady state can be characterized as \cite{Krug91}: i) low density (LD) phase with average density $\langle\rho\rangle=\alpha$, ii) high density (HD) phase $\langle\rho\rangle=1-\beta$, and iii) a maximal current (MC) phase $\langle\rho\rangle=1/2$.  Along $\alpha=\beta<1/2$, LD and HD coexist with a sharp ``shock'' wandering throughout the system.  The coexistence line is referred to as the shock phase (SP).  The exact steady state solution is known \cite{Derrida92,Derrida93,Schutz93} (for a recent review see \cite{Blythe07}).

In most transport systems, however, the hopping rate $\gamma$ of particles is rarely homogeneous throughout the system. The supply of particles can also be far from thermodynamic limit, sometimes even on par with the system size.  The interplay of inhomogeneous hopping and finite reservoir regulates the overall flux, a quantity that characterizes the steady state of the entire system.  We note that despite much effort in studying variations of TASEP, the effect due to local defect in the presence of limited particle supply remains unexamined. Our study here reveals a regime where the steady state current is limited by the strength of the local defect despite tuning of the entry. Utilizing a domain wall approach, we also identify the localization of the shock demonstrated in the density profiles in this regime. 

We provide a brief summary of relevant earlier studies here.  More details on how either aspect affects the system properties respectively are reviewed in the next section.

Open TASEP with a single defect restricted in the middle of the lattice was first studied in \cite{Kolomeisky98}.  Using a mean-field approach, Refs. \cite{Kolomeisky98} and \cite{Ha03} provided approximations for the bulk densities on either side of the defect as well as the current in steady state.  In \cite{Ha03}, scaling exponents are found numerically for the deviations of the density profile near the defect. Driven by both mathematical and biological communities, multiple slow sites, including fully inhomogeneous cases, have been under investigation \cite{Chou04,Dong07b,Greulich08,Shaw04,Pierobon06, Foulaadvand08,Harris04,Zia11}. Even though a complete analytical picture for TASEP with inhomogeneity remains elusive, various levels of mean-field approximations proved successful in capturing the current \cite{Shaw04, Harris04}, a key quantity characterizing the steady state of the system. Additionally, TASEP with extended particles \cite{Shaw04b, Dong07b, Dong07} and recharge dynamics \cite{Brackley10b,Brackley10,Turci13} have been studied.

On the front of finite particle reservoir\cite{Adams08, Cook09, Cook09b, Greulich12, Cook12}, various rules have been implemented to regulate particles' entry to the system. For instance, \cite{Ha02} investigated the parking garage problem and applied a constant entry rate until the particle pool is depleted. In the context of translocating ribosomes on a messenger RNA template\cite{Adams08}, the entry rate depends on the number of particles remaining in the pool. 
In the simple case where a reservoir of ribosomes supplies to a single mRNA, a crossover regime appears as the system transitions from LD to HD \cite{Adams08}, which is characterized by a localized ``shock'' in the density profile  \cite{Cook09}.  Refs.\ \cite{Cook09b, Greulich12, Cook12} presented more generalizations on multiple mRNAs competing for ribosomes from the same reservoir and identified the relation between lattice parameters and the reservoir density. 

When the steady state flux is in question, the regulation from entry and hopping rate of particles can affect the system in various ways. In the case of ribosomes translating on mRNA, optimizing the overall current explores the general regime between entry-limiting and hopping-limiting scenarios. Similarly in regulating traffic flow, a slow segment on the road may well cancel the efforts in tuning the ``on-ramp.''

To confirm these intuitions and chart the entry/exit-limiting and hopping-limiting regions, we study the effects of TASEP coupled to a finite pool of particles with a defect located at its center in this article. Through several refined mean-field approximations, we explicate on how the defect site and the regulated entrance from a finite pool of particles affect the overall average density, current, and density profiles in a wide range of parameters.  We see novel behaviors emerging in these quantities, particularly in density profiles.  The paper is organized as follows:  we first review relevant results in section \ref{Section2}.  We define our model and present our simulation results for a single TASEP in section \ref{Section3}.  Several analytical approaches are presented in section \ref{Section4}.  Finally, we summarize and provide further avenues of investigation in section \ref{Section5}.

\section{Synopsis of previous results}\label{Section2}

\subsection{Finite resources}

Previous studies of the open TASEP with finite resources include both single \cite{Ha02,Adams08,Cook09} and multiple TASEPs \cite{Cook09b,Greulich12,Cook12}.  In these studies, the entry rate depends on the number of particles remaining in the pool $N_p$ through the expression
\begin{equation}\label{a-eff}
\alpha_{\text{eff}}=\alpha\tanh\left(\frac{N_p}{N^*}\right)
\end{equation}
where $\alpha$ is the entry rate with unlimited resources and $N^*$ is a scaling factor that controls the strength of the feedback. When $N_p \ll N^\ast$, indicating a scarce supply of particles, then the entrance grows linearly with the pool. When more particles are added to the reservoir, $N_p \gtrsim N^\ast$, the entry rate returns to the unlimited resources situation. Since the particles are recycled back into the pool once they leave the TASEP, the total number of particles $N_{\text{tot}}$ in the system (pool + TASEP) remains constant.  For the single TASEP case, the three phases from the original TASEP reappear \cite{Adams08}.  However, a new type of SP appears when the system crosses from LD to HD.  The shock is localized to a small portion of the lattice for a range of $N_{\text{tot}}$ values \cite{Cook09}.  The strength of the localization depends on $N^*$ and the number of TASEPs utilizing the particles in the pool \cite{Cook09}.  When  the pool is supplying several TASEPs, various combinations of the LD, HD, MC, and SP (with and without shock localization) can occur \cite{Cook09b,Greulich12,Cook12}.

While mean-field results capture most aspects of the current and average density \cite{Adams08}, we need to use a domain wall (DW) approach \cite{Kolomeisky98b} to understand the features in the density profiles manifested in Monte Carlo simulations \cite{Cook09}.  On the lattice, a LD region and a HD region are separated by a sharp DW.  This DW performs a random walk along the lattice of length $L$ and is being reflected back into the lattice at the boundaries.  With a finite pool of particles, the motion of the DW changes the number of particles remaining in the pool, which in turn affects the value of $\alpha_{\text{eff}}$.  By rewriting $N_p$ in terms of the domain wall position $x$, $\alpha_{\text{eff}}$ now explicitly depends on the site that the DW is located \cite{Cook09}
\begin{equation}
\alpha_{\text{eff},x}=\alpha\tanh\left(\frac{N_{\text{tot}}-(1-\beta)(L-x)-x\alpha_{\text{eff},x}}{N^*}\right)
\end{equation}
This self-consistent equation allows us to find $\alpha_{\text{eff},x}$ numerically for each value of $x$.  By setting the density of LD and HD regions to be $\alpha_{\text{eff},x}$ and $1-\beta$, respectively, the hopping rates at site $x$ become site-dependent \cite{Cook09}
\begin{align}
D^+_x&=\frac{\beta(1-\beta)}{1-\beta-\alpha_{\text{eff},x}}\\
D^-_x&=\frac{\alpha_{\text{eff},x}(1-\alpha_{\text{eff},x})}{1-\beta-\alpha_{\text{eff},x}}
\end{align}
The probability of finding the DW at site $x$, $P(x)$, can be obtained from the master equation for the motion of the DW.  The density profile is calculated from the $P(x)$ distribution \cite{Cook09}.  Additionally, the DW approach accurately predicts the $N_{\text{tot}}$ value at which the system transitions from LD to the new SP and from SP to HD.

\subsection{Defects}

Both a naive mean-field (NMF) approach \cite{Kolomeisky98,Ha03} and finite segment mean-field theory (FSMF) \cite{Chou04} have been employed to study open TASEPs with defects. In the presence of a single defect with hopping rate larger than $\gamma=1$, the density profile remains the same as the ordinary TASEP except in a localized region around the defect.  For a defect with hopping rate $q<1$, the overall system is found to be either $q$-limiting or $\alpha$ (or $\beta$)-limiting. In the former, particles pile up behind the defect.  The density profile contains a HD region jointed through the defect to a LD region.  Neglecting spatial correlations across the slow site, \cite{Kolomeisky98} computed the densities for the two regions as well as the steady state current:
\begin{align}\label{SMF-densities}
\rho_H=&\frac{1}{1+q}\\
\rho_L=&\frac{q}{1+q}\\
J=&\rho_{H/L}(1-\rho_{H/L})=\frac{q}{(1+q)^2}
\end{align}
In the case of $\alpha$ (or $\beta$)-limiting, the system returns to an ordinary TASEP found either in LD or HD with a continuous density profile.

For particular values of $q$, spatial correlations become important in determining the current \cite{Chou04,Dong07,Greulich08,Ha03}.  To account for the correlations, FSMF computes the exact current of an $n$-site lattice segment including the defect and matches it to the currents in the bulk of the two sub-lattices \cite{Chou04}.  This approach is performed numerically up to $n\sim 20$. Comparing the results with Monte Carlo simulations, the authors found that typically only three or four sites are needed to obtain accurate results.

\section{Model definition and simulation results}\label{Section3}

Our model consists of a single open TASEP of length $L$ with a slow site located in the center of the lattice.  Meanwhile, the system is coupled to a finite pool of particles.  Particles enter the system from the pool (provided the first site is empty) with probability $\alpha_{\text{eff}}$ given in equation \eqref{a-eff}.  Each particle on the lattice moves to the adjacent vacant site with unit probability, except at the defect with probability $q<1$.  Particles leave the system at the last site of the lattice with probability $\beta$, and is immediately recycled back into the pool of particles.  The pool and lattice sites are updated sequentially at random.  We define a Monte Carlo step (MCS) to be $L+1$ updates, giving equal probability to update each lattice site as well as the reservoir.  We typically discard the first $10^5$ MCS to allow the system to reach steady-state.  We then run our simulations for an additional $10^6$ MCS, taking data every 100 MCS.  Measurements are averaged over 100 such runs.

In our model, we focus on the interplay among the following parameters : $\alpha$, $\beta$, $q$, $L$, and $N_\text{tot}$. In its simplest version, the slow site is fixed in the middle of the system. We also keep the scaling factor $N^*=750$ for all of our simulations. It is worth noting that both the location of the slow site and the strength of the feedback (determined by $N^*$) play non-trivial roles in the steady state properties of the system. We will save the investigation on these effects for future studies. With the remaining parameters, three scenarios emerge to characterize the current: entry-limiting case, exit-limiting case, and defect-limiting case.  With the system coupled to a finite reservoir, the entry-exit symmetry in an ordinary TASEP does not survive. Instead, we discover a new phenomenon in the defect limiting case to be detailed below.

\subsection{Defect-limiting case}

\begin{figure}[tbh]
\begin{center}
\includegraphics[width=8.6cm]{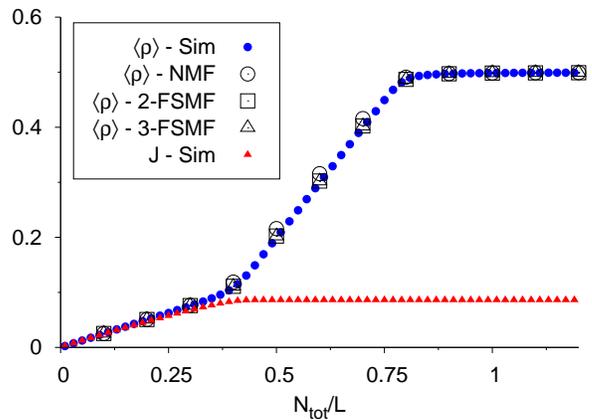}
\caption{(Color online) Overall (blue circles) average density and (red triangles) current as a function of $N_{\text{tot}}/L$ with $\alpha=0.25$, $\beta=0.75$, and $q=0.1$.\label{sim-LSP-density-current}}
\end{center}
\end{figure}

The defect-limiting case presents the most qualitatively different insights from previous studies.  In this case, we have $\rho_L<\alpha$ and $\rho_L<\beta$.   Figure \ref{sim-LSP-density-current} (with $\alpha=0.25$, $\beta=0.75$, and $q=0.1$) illustrates the behavior of the overall average density $\langle\rho\rangle$ and current $J$.  
Three different regimes form as the $N_{\text{tot}}$ increases.  The first regime occurs when $N_{\text{tot}}$ is small.  In this regime, the system lacks particles to fill the lattice, resulting in a LD state.  As $N_{\text{tot}}$ increases, the density rises linearly due to the linear nature of $\alpha_{\text{eff}}$ for small pools.  Once $\alpha_{\text{eff}}=\rho_L$, the system enters the second regime.  This regime is marked by $\alpha_{\text{eff}}$ remaining constant as $N_{\text{tot}}$ continues to increase. Also, the current has already reached its limit at the beginning of the second regime, yet the density increases linearly. Finally, the system becomes saturated at $\langle\rho\rangle=0.5$ in a MC state for the third regime.

\begin{figure}[tbh]
\begin{center}
\includegraphics[width=8.6cm]{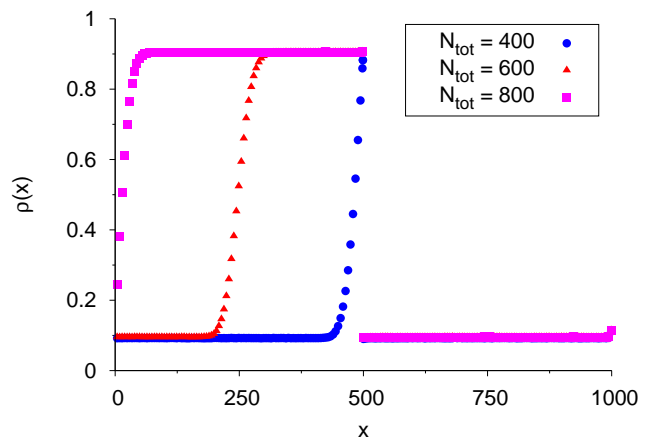}
\caption{(Color online) Density profile with $\alpha=0.25$, $\beta=0.75$, and $q=0.1$ for (blue circles) $N_{\text{tot}}=400$, (red triangles) $N_{\text{tot}}=600$, and (purple squares) $N_{\text{tot}}=800$.\label{sim-LSP-profile}}
\end{center}
\end{figure}

During the second regime, a localized shock emerges from the defect and moves toward the entrance as $N_{\text{tot}}$ increases.  An example of this movement is shown in figure \ref{sim-LSP-profile} for $\alpha=0.25$, $\beta=0.75$, and $q=0.1$. 
The formation of the localized shock at the defect site differs from previous results \cite{Cook09} where shock formed at the exiting site. The shock on the sub-lattice left of the defect separates a HD region $\rho_H$ and a LD one $\rho_L$.  The rate $q$ determines the values of $\rho_H$ and $\rho_L$, but the values seen in the simulation differ from what is estimated using equation \eqref{SMF-densities}.  Further, the location about which the shock is localized depends on both $\rho_H$ and $\rho_L$.  However, the site where the shock is localized does not alter the density profile to the right of the defect.

\subsection{Boundary-limiting cases}

When the current is controlled by the boundary rates, the system behaves similar to a constrained TASEP.  For completeness, we present the simulation results for the entry-limiting and exit-limiting cases, noting the differences from the constrained TASEP.  When $\alpha<\rho_L$ and $\alpha<\beta$, the entry rate of particles limits the current in the system.  This situation corresponds to the LD phase in the ordinary TASEP.  Particles will not enter the lattice at a fast enough rate to be held up by the defect or exit.  Therefore, $\langle\rho\rangle$ and $J$ are controlled by the value of $\alpha_{\text{eff}}$.  This behavior is reminiscent of a constrained TASEP \cite{Adams08} for $\alpha<\beta$.  While the defect site has little effect on $\langle\rho\rangle$, it has more influence on the density profile.

\begin{figure}[tbh]
\begin{center}
\includegraphics[width=8.6cm]{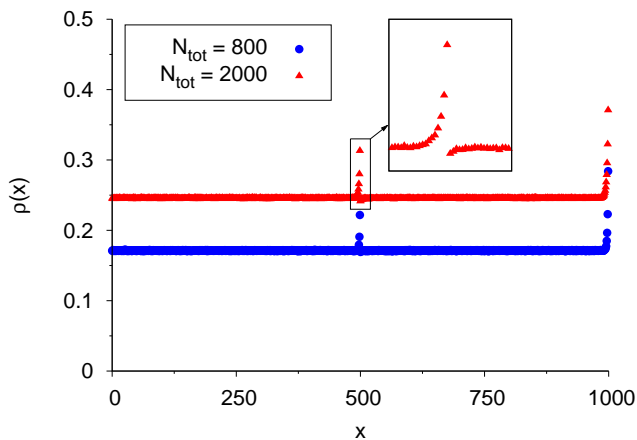}
\caption{(Color online) Density profile with $\alpha=0.25$, $\beta=0.5$, and $q=0.75$ for (blue circles) $N_{\text{tot}}=800$ and (red triangles) $N_{\text{tot}}=2000$.  Inset shows the deviation from the bulk density near the defect.\label{sim-LD-profile}}
\end{center}
\end{figure}

While the defect does not receive particles fast enough to create a large back up, it does hold up a small amount of particles just longer than other sites on average.  As shown in figure \ref{sim-LD-profile}, this creates an increase in the density profile above the bulk density just before the defect and a decrease (below the bulk density) immediately after it.  
A similar feature appears in an ordinary TASEP with a defect \cite{Kolomeisky98,Ha03}.  These ``kicks'' only affect the site density of a few sites and are not symmetric about the defect.  The drop below the bulk density in the profile after the defect is much less pronounced than the increase before the site as seen in the inset of fig \ref{sim-LD-profile}.  A naive mean-field approach fails to capture this subtle feature as significant spatial correlations are introduced in the presence of the defect.

\begin{figure}[tbh]
\begin{center}
\includegraphics[width=8.6cm]{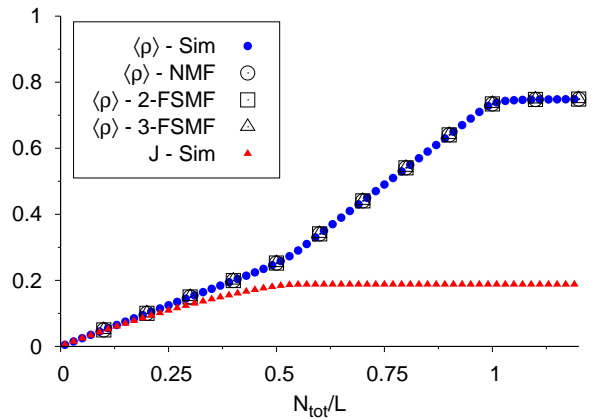}
\caption{(Color online) Overall (blue circles) average density and (red triangles) current as a function of $N_{\text{tot}}/L$ with $\alpha=0.75$, $\beta=0.25$, and $q=0.5$.\label{sim-HD2-density-current}}
\end{center}
\end{figure}

When $\beta<\alpha<\rho_L$, the exit rate controls the current through the system with large $N_{\text{tot}}$.  This case is analogous to the HD phase in the ordinary TASEP.  The number of particles and boundary rates play an important role in controlling the overall density and current, while the defect has little effect.  An example of how $\langle\rho\rangle$ and $J$ change with $N_{\text{tot}}$ is shown in figure \ref{sim-HD2-density-current} for $\alpha=0.75$, $\beta=0.25$, and $q=0.5$. 
Again, this case is similar to what is seen in the constrained TASEP \cite{Adams08} with $\beta<\alpha$.  Three regimes exist in figure \ref{sim-HD2-density-current}.  The second regime corresponds to the system crossing the LD-HD phase boundary.  The density rises linearly (but with a unit slope) while the current remains constant.  Also in this regime, the average number of particles in the pool remains constant as $N_{\text{tot}}$ increases.

\begin{figure}[tb]
\begin{center}
\includegraphics[width=8.6cm]{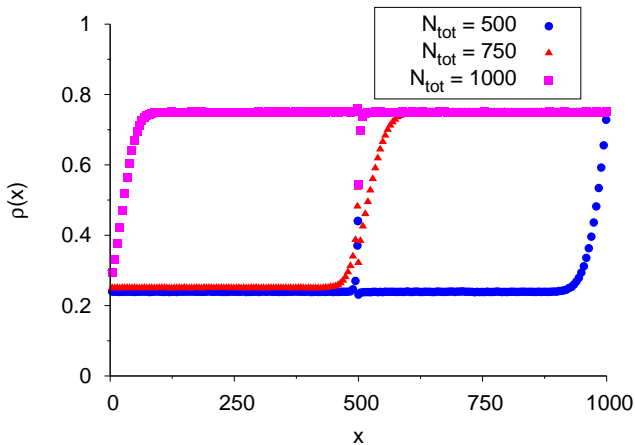}
\caption{(Color online) Density profile with $\alpha=0.75$, $\beta=0.25$, and $q=0.5$ for (blue circles) $N_{\text{tot}}=500$, (red triangles) $N_{\text{tot}}=750$, and (purple squares) $N_{\text{tot}}=1000$.\label{sim-HD2-profile}}
\end{center}
\end{figure}

The density profile resembles the constrained TASEP \cite{Cook09} as $N_{\text{tot}}$ varies.  As with the previous case, the slow site causes the particles to pile up behind it leading to an increase in the site density before the defect and a drop in the density afterwards.  At small $N_{\text{tot}}$ values, $\alpha_{\text{eff}}$ limits the number of particles entering the lattice; thus, the profiles are similar to the entry-limited case.  As $N_{\text{tot}}$ increases, eventually $\alpha_{\text{eff}}=\beta$ signaling the start of the second regime mentioned above.  In this regime, a localized shock appears near the exit, and moves toward the entrance as $N_{\text{tot}}$ increases as shown in figure \ref{sim-HD2-profile}.  
The relatively high $q$ value allows particles to move quickly through the middle of the lattice, creating only a local disturbance, even as the shock moves through the slow site.  Once the shock reaches the entrance, TASEP is saturated with particles, and the profile remains fixed as $N_{\text{tot}}$ increases in this HD state.

\section{Theoretical approach}\label{Section4}

Even though the steady state current $J$ behaves qualitatively similar in the aforementioned three cases, the density profiles in entry-, exit- and defect- limiting cases are quite different. To further elucidate the latter, we turn to a domain wall (DW) approach which has been successful in previous studies \cite{Cook09b, Cook12} in capturing the localization of the shock and reproduced the density profile resulting from the Monte Carlo simulations.

\subsection{Domain wall for boundary-limiting cases}

\begin{figure}[tbh]
\begin{center}
\includegraphics[width=8.6cm]{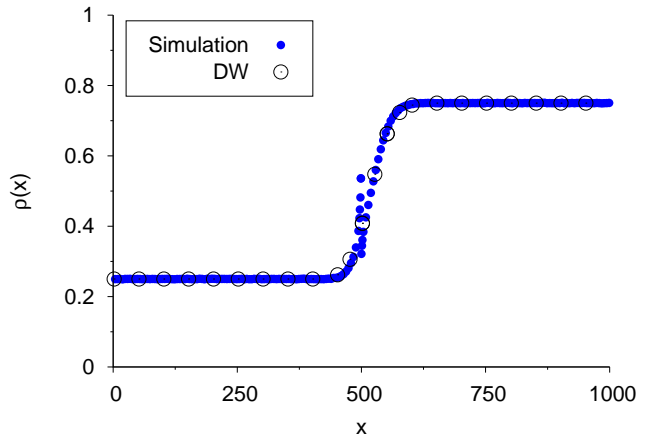}
\caption{(Color online) Comparison of the (blue circles) simulation and (open circles) domain wall results for the density profile with $\alpha=0.75$, $\beta=0.25$, and $q=0.5$ for $N_{\text{tot}}=750$.\label{theory-HD-profile}}
\end{center}
\end{figure}

Using DW results discussed in section \ref{Section2}, we have similar success for the entry- and exit- limiting cases for the overall density $\langle\rho\rangle$, shown in figure \ref{sim-HD2-density-current}. Additionally, DW captures the shock localization in the exit-limiting case, as shown in figure \ref{theory-HD-profile}, even when the localization region includes the defect. 
While the location of the shock is captured by the theory, the kicks around the defect are not.  These kicks are due to the spatial correlations induced by the defect; yet, DW approach only uses the boundary rates as input.  DW will not capture this detail of the profile.  However, we can conclude for this case that the defect has no impact on the location of the shock localization, which only depends on the entry and exit rates.

\subsection{Domain wall for defect-limiting case}

To better understand the defect-limiting case, we look at how $q$ affects the density profile for either sub-lattice.  As seen in the simulation, the defect produces strong spatial correlations within a few sites.  A naive mean field (NMF) approach, which ignores such correlations, provides a reasonable estimate for $\langle\rho\rangle$ and $J$.  However, the lack of information about spatial correlations makes these results inadequate for producing the density profiles \cite{Kolomeisky98, Ha03}, even with DW approach.  To properly account for the correlations around the defect, we employ the finite segment mean field (FSMF) theory used in Ref. \cite{Chou04} to compute densities for the LD and HD regions as additional inputs for DW theory.  Since the kicks near the slow site decay into the bulk density very quickly, only a few sites are needed for the FSMF to give descent estimates of the bulk density.

We calculate the bulk densities to the left and right of the slow site, denoting them as $\rho_H$(to the left) and $\rho_L$ (to the right).  For an $n$-site FSMF, the defect is positioned at $n/2$ for even $n$, or $(n+1)/2$ for odd.  Setting up the $2^n\times 2^n$ transition matrix between configurations, we solve for the eigenvector that corresponds to the zero-eigenvalue.  We then calculate the current $J$ using the eigenvector, and match it to the boundary currents $J=\rho_L(1-\rho_L)=\rho_H(1-\rho_H)$.  Clearly, $\rho_H=1-\rho_L$ is the only physical solution for the boundaries.  This leads to at least an $n$-order polynomial in $\rho_H$. The two-site polynomial is
\begin{equation}
0=2\rho_H^2+\rho_H(3q-2)-q
\end{equation}
and for three sites, we obtain
\begin{multline}
0=4\rho_H^4(1+q)+4\rho_H^3(q+q^2)\\
+\rho_H^2(3q^2+3q-4)+\rho_H(q^2-3q)-q^2
\end{multline}

\begin{figure}[tbh]
\begin{center}
\includegraphics[width=8.6cm]{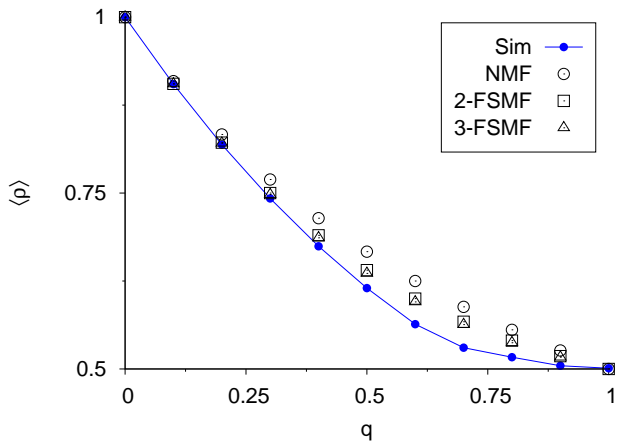}
\caption{(Color online) Bulk density for the left sub-lattice as a function of $q$ for various mean-field approaches.  The simulation curve keeps the entry and exit rates constant with $\alpha=\beta=1$ for $L=1000$.\label{Q-Density-compare}}
\end{center}
\end{figure}

The numerical solutions for the densities from FSMF theory give an improvement over NMF theory when compared to the simulation data as shown in figure \ref{Q-Density-compare}.
The simulation densities are the average site densities of 100 sites away from the slow site and boundaries.  Additionally, $\alpha$ and $\beta$ are kept constant and set to unity.  The two-site FSMF (2-FSMF) is a major improvement over NMF, while the three-site FSMF (3-FSMF) gives a slightly better result than the two site.  Increasing $n$ would, in principle, improve the FSMF results.  But the improvement in predicted $J$ and $\rho$ is already quite minimal when $n=3$, as shown in figure \ref{Q-Density-compare}, a trend consistent with the findings in \cite{Chou04}.  For the results that we present in this paper, we include up to the 3-FSMF for comparison with the simulation data.

The left and right bulk densities computed from FSMF are incorporated into the DW theory.  The effect of the defect site on the bulk densities only becomes apparent when the system is defect-limiting, which we will focus on.  In this case, we treat the left and right sub-lattices as having separate domain wall dynamics.  For the left sub-lattice, the slow site becomes an effective boundary and controls the hopping rates of the domain wall on it.  $\rho_H$ from the FSMF is used to determine the hopping rates of the shock.  Thus, we have for the hopping rates
\begin{eqnarray}
D^+_x&=&\frac{\rho_H(1-\rho_H)}{\rho_H-\alpha_{\text{eff},x}}\\
D^-_x&=&\frac{\alpha_{\text{eff},x}(1-\alpha_{\text{eff},x})}{\rho_H-\alpha_{\text{eff},x}}
\end{eqnarray}
where $\alpha_{\text{eff},x}$ is calculated from the following self-consistent equation:
\begin{equation}
\alpha_{\text{eff},x}=\alpha\tanh\left(\frac{N_{\text{tot}}-\rho_L(L-k)-\rho_H(k-x)-x\alpha_{\text{eff},x}}{N^*}\right)
\end{equation}
Same as the case without a slow site \cite{Cook09}, the hopping rates have a site dependence due to the fluctuating value of $\alpha_{\text{eff}}$.

The DW performs a biased random walk throughout the left sub-lattice with reflecting boundary conditions at the entrance and site $k$, of which the motion is governed by the following master equation:
\begin{equation}\label{master-single}
\frac{\partial P(x)}{\partial t}=0=D^+_{x-1}P(x-1)+D^-_{x+1}P(x+1)-(D^+_x+D^-_x)P(x)
\end{equation}
The boundary conditions are given by:
\begin{eqnarray}
\frac{\partial P(0)}{\partial t}&=0=&D^-_1P(1)-D^+_0P(0)\\
\frac{\partial P(k)}{\partial t}&=0=&D^+_{k-1}P(k-1)-D^-_kP(k)
\end{eqnarray}
We find the solution to equation \eqref{master-single}:
\begin{equation}
P(x)=
\begin{cases}
\dfrac{1}{Z} & x=k\\
\dfrac{1}{Z}{\displaystyle\prod_{j=x}^{k-1}}\dfrac{D^-_{j+1}}{D^+_j} & 0\le x\le k-1
\end{cases}
\end{equation}
where $Z$ is the normalization constant:
\begin{equation}
Z=1+\sum_{x=0}^{k-1}\prod_{j=x}^{k-1}\frac{D^-_{j+1}}{D^+_j}
\end{equation}
Subsequently, we obtain the density profile $\rho_x$ using
\begin{equation}
\rho_x=\sum_{i=0}^x\rho_H P(i)+\sum_{i=x+1}^k\alpha_{\text{eff},i}P(i)
\end{equation}
for the left sub-lattice and $\rho_x=\rho_L$ for the right sub-lattice. Averaging $\rho_x$ over the entire lattice gives us the overall average density $\langle\rho\rangle$.

\begin{figure}[tbh]
\begin{center}
\includegraphics[width=8.6cm]{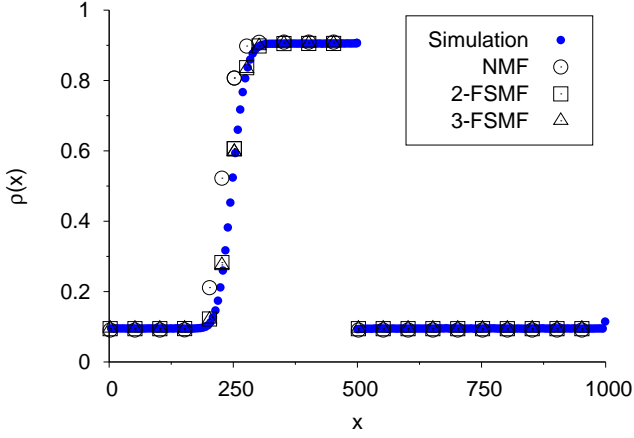}
\caption{(Color online) Comparison of density profiles between (blue circles) simulation and (open points) FSMF DW theory with $\alpha=0.25$, $\beta=0.75$, and $q=0.1$ for $N_{\text{tot}}=600$.\label{theory-compare-q10}}
\end{center}
\end{figure}

The agreement between DW theory and the simulation data for the slow site limited case depends on the accurately predicting the bulk densities from $q$.  For $q$ near 0 or 1, both NMF and FSMF approximate the densities seen in the simulations.  Figures \ref{sim-LSP-density-current} and \ref{theory-compare-q10} show the comparison between using NMF, 2-FSMF, and 3-FSMF for $\rho_H$ in the DW theory with $q=0.1$.
The difference in both figures is minimal:  the 2-FSMF and 3-FSMF results provide only a slight improvement over NMF.  However, NMF result for $\rho_H$ with $q=0.1$ is very close to the simulation results, as shown in figure \ref{Q-Density-compare}.  Thus, our DW approach correctly predicts the density profile and overall density when we first compute an accurate value for the bulk densities.

\begin{figure}[tbh]
\begin{center}
\includegraphics[width=8.6cm]{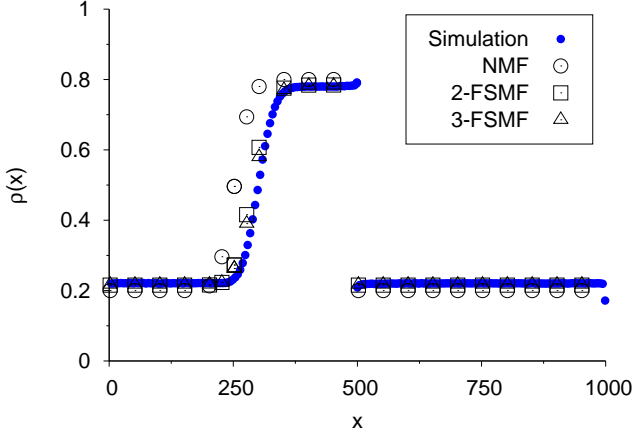}
\caption{(Color online) Comparison of density profiles between (blue circles) simulation and (open points) FSMF DW theory with $\alpha=\beta=1$, and $q=0.25$ for $N_{\text{tot}}=500$.\label{theory-compare-q25}}
\end{center}
\end{figure}

\begin{figure}[tbh]
\begin{center}
\includegraphics[width=8.6cm]{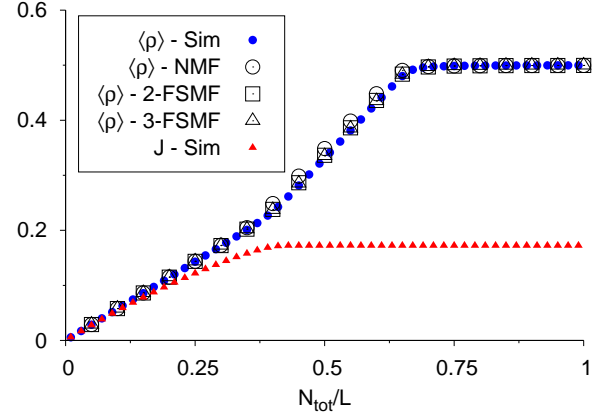}
\caption{(Color online) Overall (blue circles) average density and (red triangles) current as a function of $N_{\text{tot}}/L$ for simulations and (open points) FSMF DW theory with $\alpha=\beta=1$ and $q=0.25$.}
\label{theory-compare-density-q25}
\end{center}
\end{figure}

To show how the bulk densities affect the DW predictions, we look at $q$ away from 0 and 1.  For $q=0.25$, the difference is noticeable in the density profile as seen in figure \ref{theory-compare-q25}.
The 2-FSMF and 3-FSMF results are in much better agreement with the simulation results than NMF; however, all three results predict a greater $\rho_H$ than the simulations.  $\rho_H$, which appears in both $D^+_x$ and $D^-_x$, controls both the height of the DW and the location of where the shock is localized.  The 2- and 3-FSMF give more accurate overall average density value, but the difference with the NMF is less pronounced (figure \ref{theory-compare-density-q25}).

\subsection{LD-MC phase transition}

\begin{figure}[tbh]
\begin{center}
\includegraphics[width=8.6cm]{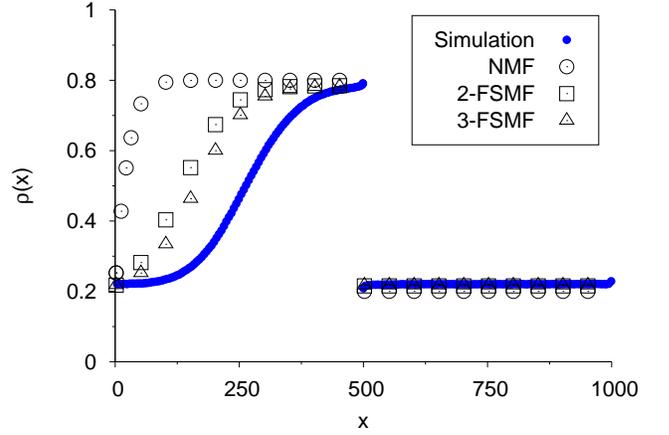}
\caption{(Color online) Comparison of density profiles between  (blue circles) simulation and (open points) FSMF DW theory with $\alpha=q=0.25$ and $\beta=0.75$ for $N_{\text{tot}}=1400$.\label{theory-FSMFT-compare}}
\end{center}
\end{figure}

The phase transition between the LD and MC phases occurs when the average $\alpha_{\text{eff}}=\rho_L$.  The difference in the 2-FSMF and 3-FSMF results appears when the system is near this transition line for large $N_{\text{tot}}$.  In figure \ref{theory-FSMFT-compare}, we have $\alpha=q=0.25$ and $\beta=0.75$.  
As seen in figure \ref{theory-FSMFT-compare}, 3-FSMF result better captures our simulation than 2-FSMF; however, neither gives an accurate prediction of the density profile.  Similar disagreement exists for $\langle\rho\rangle$.  The inclusion of more sites in the FSMF continues to improve the agreement between the DW theory and the simulations.  The need for more sites in the FSMF is a result of the strong spatial correlations between lattice sites near the LD-MC phase boundary.

\begin{figure}[tb]
\begin{center}
\includegraphics[width=8.6cm]{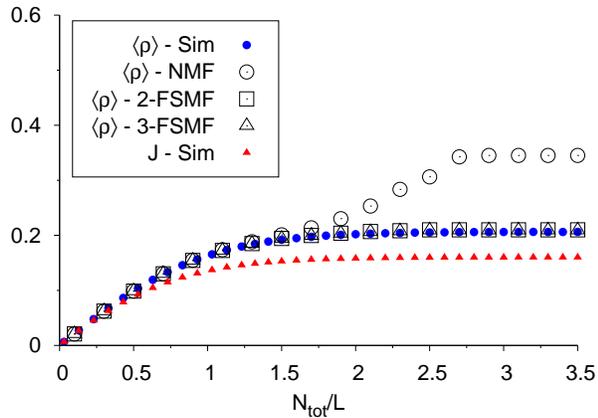}
\caption{(Color online) Overall (blue circles) average density and (red triangles) current as a function of $N_{\text{tot}}/L$ for simulations and (open points) FSMF DW theory with $\alpha=0.2$, $\beta=0.75$ and $q=0.25$.}
\label{theory-phase-transitions}
\end{center}
\end{figure}

In figure \ref{theory-phase-transitions}, we see a failure of NMF to correctly predict $\langle\rho\rangle$ near the LD-MC phase boundary for large $N_{\text{tot}}$.  NMF theory predicts that the system will leave LD, while simulations show that it remains in LD.  The system begins transitioning from LD to MC when $\langle\rho\rangle=\rho_L$ which occurs when the average $\alpha_{\text{eff}}$ reaches $\rho_L$:
\begin{equation}\label{low-Ntot}
\rho_L=\alpha\tanh\left(\frac{N_{\text{tot}}-\rho_LL}{N^*}\right)
\end{equation}
and enters MC when the average density on the left sub-lattice is $\rho_H$
\begin{equation}\label{high-Ntot}
\rho_L=\alpha\tanh\left(\frac{N_{\text{tot}}-L/2}{N^*}\right)
\end{equation}
For parameters shown in figure \ref{theory-phase-transitions}, NMF predicts that the left sub-lattice should have a delocalized DW at large $N_{\text{tot}}$ values for $\alpha=0.2$ and $q=0.25$, which lead to $\langle\rho\rangle>\alpha$.  Yet, the simulation results reveal that the system remains in the LD phase for very large $N_{\text{tot}}$ values.  
The FSMF results have better agreement with what is seen in the Monte Carlo simulations as shown in figure \ref{theory-phase-transitions}; however, these will also fail as $\alpha$ continues to approach $\rho_L$.

\section{Summary and Outlook}\label{Section5}

Motivated by transport processes with limited resources as well as ``speed bumps,'' we investigate TASEP coupled with a finite supply of particles and defect site in the middle of the system. The reservoir of the particles dictates the entry rate through equation \eqref{a-eff} and the parameter $q$ regulates the strength of the defect. By means of a combination of Monte Carlo simulations and various levels of mean field approximations, we presented the results and a quantitative understanding of the interplay amongst limited resources and a single defect.  We demonstrated that a domain wall theory, which uses the defect as an effective boundary, captures the localization of the shock seen in the simulations.  Although it failed to provide accurate quantitative results when the system was near the LD-MC phase transition for large $N_{\text{tot}}$, it provided valuable qualitative insight in this regime.

Beyond the study presented in this article, several open questions remain intriguing.  While we have limited our study to a defect located in the center, this does not need to be the case.  The location of the slow site affects the density and current in the simplest TASEP \cite{Dong07b}.  Thus, a natural extension of our study is to explore the affects of the slow site's location in a TASEP with finite resources.  Returning to our motivation from biology, ribosomes cover more than one codon.  Having large particles covering more than one site in our model would be more realistic.  Also, many mRNA compete for the same resources, so what effect does a slow site have on competition between TASEPs coupled to the same pool of particles?  Finally, we still do not have an analytical way of predicting the bulk densities, which should be further studied in the future.  Transport phenomena play a vital role in many system in nature.  Our study has furthered the understanding of such systems by exploring the coupling between finite resources and defects.  However, search for a comprehensive framework for solving these systems is still underway.  Many more studies must be performed in the future before we can say that we truly understand the world around us.

\begin{acknowledgments}
The authors would like to thank Royce K. P. Zia for his illuminating discussions.  We would also like to thank the anonymous referees who provided additional reference and useful comments.  LJC is supported by the Lenfest Summer Research Grant.  JJD is supported by NSF grant DMR-1248387.  ALF is supported by Robert E.\  Lee Summer Scholars Program.
\end{acknowledgments}

\bibliography{references}

\end{document}